\documentclass[preprint,preprintnumbers, prd, floatfix, superscriptaddress,nofootinbib] {revtex4-1}
\usepackage{epsfig}
\usepackage{subfigure}
\usepackage{dcolumn}
\usepackage{bm}
\usepackage[usenames ,dvipsnames]{xcolor}
\usepackage{slashed}
\usepackage{graphicx,color}
\usepackage[bookmarksnumbered, pdfstartview=FitH,colorlinks,urlcolor=blue, citecolor=blue,linkcolor=blue] {hyperref}
\usepackage{appendix}
\usepackage{tabularx}
\begin{document}
\title{Five-body $D\to V$ Semileptonic Decays}
\author{Yechun Yu}
\affiliation{School of Physics, Zhengzhou University, Zhengzhou, Henan 450001, China}
\author{Han Zhang}
\email{Corresponding author: zhanghanzzu@gs.zzu.edu.cn}
\affiliation{School of Physics, Zhengzhou University, Zhengzhou, Henan 450001, China}
\author{Bai-Cian Ke}
\email{Corresponding author: baiciank@ihep.ac.cn}
\affiliation{School of Physics, Zhengzhou University, Zhengzhou, Henan 450001, China}
\author{Yao Yu}
\email{Corresponding author: yuyao@cqupt.edu.cn}
\affiliation{Chongqing University of Posts \& Telecommunications, Chongqing, 400065, China}
\affiliation{Department of Physics and Chongqing Key Laboratory for Strongly Coupled Physics, Chongqing University, Chongqing 401331, People's Republic of China}
\author{Zhuang Xiong}
\affiliation{Chongqing University of Posts \& Telecommunications, Chongqing, 400065, China}
\author{Jia-Wei Zhang}
\affiliation{Department of Physics, Chongqing University of Science and Technology, Chongqing, 401331,
 China}
 \author{Xue-Wen Chen}
\affiliation{Department of Physics, Chongqing University of Science and Technology, Chongqing, 401331,
 China}
\begin{abstract}
Our main objective is to derive the decay rate for the semileptonic decays
$D\to V\ell^+\nu_{\ell}\,(\ell=e,\mu)$, where $V$ represents a vector particle.
In these decays, the vector particle $V$ decays into three pseudo-scalar
particles. To accomplish this, we evaluate the phase-space factor for the
five-body decay with a set of eight independent variables which uniquely define
a point in the phase space. We further conduct a detailed investigation of the
$D\to \omega\ell^+\nu_{\ell}$, where $\omega$ subsequently decays into
$\pi^+\pi^-\pi^0$, within the Standard Model and in a general effective field
theory description of the weak interactions at low energies. The outcomes of
this study have potential applications in the measurement of $D\to \omega$ form
factors. These measurements can be performed using data obtained from BESIII.
\end{abstract}
\maketitle
\section{Introduction}
Semileptonic decays offer a distinct advantage in disentangling the hadronic
and weak currents, enabling a more accurate examination of strong and weak
interactions. This unique feature positions them as the primary experimental
approach for extracting the Cabibbo-Kobayashi-Maskawa (CKM) matrix elements
and analyzing hadronic form factors. The hadronic form factor characterizes
the strong interaction between the initial and final hadrons, providing crucial
information about the dynamic behavior of the strong interaction during the
decay process~\cite{Ke:2023qzc}.

Focusing on the $D\to V$ form factors (where $V$ denote vector mesons),
extensive amplitude analyses of four-body semileptonic decays of charmed
mesons have been conducted. The $D\to \bar{K}^{*}(892)$ form factors
have been studied in the  $D\to \bar{K}\pi e^+\nu_e$ decay by the
CLEO, BABAR, and BESIII
collaborations~\cite{CLEO:2010enr, BaBar:2010vmf, BESIII:2015hty, BESIII:2018jjm}.
The $D\to\rho$ form factors have also been investigated by BESIII and CLEO in
the $D\to \pi\pi e^+\nu_e$ decay~\cite{CLEO:2011ab, BESIII:2018qmf}.
In addition to the above measurements, the form factors for $D\to\omega$ has
been reported in the study of the five-body decay
$D^{+}\to \pi^+\pi^-\pi^0 e^+\nu_e$ by BESIII~\cite{BESIII:2015kin}. While
five-body decays pose more challenges, they offer the advantage of introducing
three additional degrees of freedom compared to four-body
decays~\cite{Wang:2019wee}. This allows for a more thorough investigation into
the decay mechanism. However, BESIII adopted a parametrization specifically
designed for a four-body decay to analyze this five-body decay. This adoption
raises concerns as its validity was not confirmed and loses the ability to
investigate part of the angular distribution information, limiting the ability
to achieve a comprehensive understanding of the underlying mechanism.

The phase space treatment for multi-body decays is typically based on the
invariant mass approach, making it challenging to analyze spatial distribution
asymmetries. Therefore, expressing the phase space in the form of angular
distributions has a wide range of applications. The treatment of phase space
for three- and four-body decays in angular distributions is well established,
but that for five-body decays remains relatively uncommon.

Although it is well-known how to derive the phase space for five-body decays,
deriving a clear parametrization in angular distributions is important and
necessary for experimental purposes. For example, BESIII faced a challenge due
to the lack of an available parametrization and had to adopt the
parametrization for four-body decay $D\to K\pi e \nu_e$ from
BABAR~[Eq.~16 in Ref.~\cite{BaBar:2010vmf}] in the study of
$D^+\to \omega e^+ \nu_e$ with
$\omega \to \pi^+\pi^-\pi^0$~[Eq.~2 in Ref.~\cite{BESIII:2015kin}].
Additionally, the branching fraction~($\frac{1}{3}$ or $\frac{2}{3}$) of
$K^*\to K\pi$ in the BABAR paper is based on isospin, which is not applicable
to the $\omega\to\pi^+\pi^-\pi^0$ decay.

In this work, we begin by deriving the kinematics for a five-body semileptonic
decay, where the phase space can be uniquely defined by eight independent
variables~\cite{Kumar:1969jjy, Huber:2018gii}. We choose three invariant
masses, three helicity angles, and two angles between decay planes and obtain a
clear and concise phase-space factor. Previous investigations use the angle
between the resonance decay plane normal in the rest frame of the three
final-state particles and the direction of flight of the the resonance in the
$D$ rest frame. We then provide a detailed derivation of the dynamics involved
in the $D\to V\ell^+\nu_{\ell}$ decay. Finally, we will present a summary in
the last section.

\section{Kinematics of five-body decays}\label{sec:kinematics}
The decay rate of a particle of mass $m_A$ to five bodies in its rest frame can
be written in terms of an invariant amplitude $\bar{\cal A}$
as~\cite{Flores-Tlalpa:2015vga, Arroyo-Urena:2017ihp}
\begin{eqnarray}
  \Gamma&=&\frac{(2\pi)^4}{2m_{A}}\int\prod_{i=1}^{5}\frac{d^{3}q_{i}}{16\pi^{3}E_{i}}|\bar{{\cal A}}|^2\delta^{4}(\sum_{i=1}^{5}q_{i}-Q)=\frac{1}{2^{17}\pi^{11}m_A}{\cal I}\,,
  \label{eq:Gamma}
\end{eqnarray}
where $Q$ is the four-momentum of the parent particle and $q_{1\cdots5}$ denote
the four-momenta of the five final-state particles. The $\delta$ function
$\delta^{4}(\sum_{i=1}^{5}q_{i}-Q)$ enforces the conservation of four-momentum
between the parent and the final-state particles. Note that $q_{1\cdots5}$ can
be the four-momenta of $\ell^+$, $\nu_{\ell}$, $P_1$, $P_2$, $P_3$ in the case
of $D\to (V\to P_1P_2P_3)\ell^+\nu_{\ell}$, respectively.

To facilitate the study of systems composed of two or three final-state
particles, it is more convenient to specify the individual components of
various four-momenta in the rest frames of the corresponding parent systems.
Four rest frames of reference are of particular interest: the $Q$-frame, the
rest frame of the parent particle; the $Y$-frame, representing the
center-of-mass frame of the $q_1q_2$ system; the $K$-frame, representing the
center-of-mass frame of the $q_3q_4q_5$ system; and the $Z$-frame, representing
the center-of-mass frame of the $q_4q_5$ system. Defining $Y=q_1+q_2$,
$K=q_3+q_4+q_5$ and $Z=q_4+q_5$, the $\delta$ function in Eq.~(\ref{eq:Gamma})
can be generated recursively, and ${\cal I}$ can be rewritten
as~\cite{Cabibbo:1965zzb}
\begin{eqnarray}
  {\cal I} &=& \int\prod_{i=1}^{5}\frac{d^{3}q_{i}}{E_{i}}|\bar{{\cal A}}|^2\delta^{4}(\sum_{i=1}^{5}q_{i}-Q)\nonumber\\
        &=&\int d^{4}Y d^{4}K|\bar{{\cal A}}|^2\delta^{4}(Y+K-Q)\int \frac{d^{3}q_{1}}{E_{1}}\frac{d^{3}q_{2}}{E_{2}}\delta^{4}(q_1+q_2-Y)\nonumber\\
        &\cdot&\int\frac{d^{3}q_{3}}{E_{3}}d^{4}Z\delta^{4}(q_3+Z-K)\int \frac{d^{3}q_{4}}{E_{4}}\frac{d^{3}q_{5}}{E_{5}}\delta^{4}(q_4+q_5-Z)\,.
\end{eqnarray}
The integrations over the $\delta$ functions can be carried out independently
in their respective rest frames:
\begin{eqnarray}
  \int \frac{d^{3}q_{1}}{E_{1}}\frac{d^{3}q_{2}}{E_{2}}\delta^{4}(q_1+q_2-Y) &=& \frac{|\vec{q}_{1}|}{\sqrt{Y^2}}d\Omega_{1}\nonumber\\
  \int \frac{d^{3}q_{4}}{E_{4}}\frac{d^{3}q_{5}}{E_{5}}\delta^{4}(q_4+q_5-Z) &=& \frac{|\vec{q}_{4}|}{\sqrt{Z^2}}d\Omega_{3}
\end{eqnarray}
where $\Omega_1$ ($\Omega_3$) and $\vec{q}_{1}$ ($\vec{q}_{4}$) represents the
solid angle and the three-momentum of $q_1$ ($q_{4}$) in the $Y$ ($Z$) rest
frame, respectively;
\begin{eqnarray}
\int\frac{d^{3}q_{3}}{E_{3}}d^{4}Z\delta^{4}(q_3+Z-K) &=&  d Z^2\frac{\pi|\vec{Z}|}{\sqrt{K^2}}d\cos\theta_{2}\,,
\end{eqnarray}
where $\theta_2$ and $\vec{Z}$ are the polar angle and the three-momentum of $Z$
in the $K$ rest frame, respectively;\footnote{The azimuthal angle has been
integrated over without loss of generality.\label{azimuthal}}
\begin{eqnarray}
 \int d^{4}Y d^{4}K\delta^{4}(Y+K-Q) &=&  d Y^2 d K^2\frac{\pi|\vec{K}|}{m_{A}}\,,
\end{eqnarray}
where $\vec{K}$ is the three-momentum of $K$ in the $Q$ rest
frame.\footnote{Without any loss of generality, both the polar and azimuthal
angles have been integrated out.}
Subsequently, the decay rate can be elegantly expressed as
\begin{eqnarray}
  \Gamma_{5} &=& \frac{1}{2^{20} \pi^{9}m_A^3}\int X\beta_Y\beta_Z\beta_K|\bar{{\cal A}}|^2d\Omega_{1}d\Omega_{3}d\cos\theta_{2} d K^2 d Y^2 d Z^2\,,
  \label{eq:Gamma_5}
\end{eqnarray}
where $\beta_Y=2|\vec{q}_{1}|/\sqrt{Y^2}$, $\beta_Z=2|\vec{q}_{4}|/\sqrt{Z^2}$,
$\beta_K=2|\vec{Z}|/\sqrt{K^2}$ and $X=|\vec{K}|m_A$, which can be expressed in
terms of Lorentz-invariant variables
\begin{eqnarray}
|\vec{q}_{1}| &=& \frac{\sqrt{[Y^2-(m_1+m_2)^2][Y^2-(m_1-m_2)^2]}}{2\sqrt{Y^2}}\,, \nonumber\\
|\vec{q}_{4}| &=& \frac{\sqrt{[Z^2-(m_4+m_5)^2][Z^2-(m_4-m_5)^2]}}{2\sqrt{Z^2}}\,, \nonumber\\
|\vec{Z}| &=& \frac{\sqrt{[K^2-(\sqrt{Z^2}+m_3)^2][K^2-(\sqrt{Z^2}-m_3)^2]}}{2\sqrt{K^2}}\,, \nonumber\\
|\vec{K}| &=& \frac{\sqrt{[m_A^2-(\sqrt{Y^2}+\sqrt{K^2})^2][m_A^2-(\sqrt{Y^2}-\sqrt{K^2})^2]}}{2m_A}\,.
\end{eqnarray}
The region of integration is specified by
\begin{eqnarray}
0&\leq \phi_{1,3} \leq& 2\pi\,,\nonumber\\
-1\leq&\cos\theta_{1,2,3}&\leq 1\,,\nonumber\\
  (m_4+m_5)^{2}&\leq Z^2 \leq& (\sqrt{K^2}-m_3)^{2}\,, \nonumber\\
(m_1+m_2)^{2}&\leq Y^2 \leq& (m_A-\sqrt{K^2})^{2}\,, \nonumber\\
(m_3+m_4+m_5)^{2}&\leq K^2 \leq& (m_A-m_1-m_2)^{2}\,,\nonumber\\
&{\rm or}&\nonumber\\
(m_3+m_4+m_5)^{2}&\leq K^2 \leq& (m_A-\sqrt{Y^2})^{2}\,,\nonumber\\
(m_1+m_2)^{2}&\leq Y^2 \leq& (m_A-m_3-m_4-m_5)^{2}\,.
\end{eqnarray}

The amplitude $\bar{\cal A}$ can be evaluated using helicity amplitudes, which
is involved in four rest frames. To relate different rest frames, sets of
polarization vectors, denoted as $\epsilon^\mu(m)$, are introduced as helicity
bases. These polarization vectors satisfy the orthonormality and completeness
properties given by
\begin{eqnarray}
  &&\epsilon^\dagger_{\mu}(m)\epsilon^{\mu}(n)=g_{mn}\,\,\,\,(m,n=t,1,2,0)\,,\nonumber\\
  &&\epsilon^\mu(m)\epsilon^{\dagger\nu}(n)g_{m n}=g^{\mu\nu}\,,
\label{eq:orthonormality}
\end{eqnarray}
where
$g_{m n}={\rm diag}(+, -, -, -)={\rm diag}(g_{tt}, g_{11}, g_{22}, g_{00})$.
In the $Q$ rest frame, one has
\begin{eqnarray}
  Q^\mu &=& (m_A,0,0,0)\,, \nonumber\\
  K^\mu &=&(K^0,0,0,\vec{K})\,, \nonumber\\
  Y^\mu &=&(Y^0,0,0,-\vec{K})\,.
\end{eqnarray}
The polarization vectors for the $K$ system in the $Q$ rest frame are defined
to satisfy $\kappa_{Q}^\mu(m)K_\mu$ for $m=1,2,0$ and can be written as
\begin{eqnarray}
  \kappa_{Q}^\mu(t) &=& \frac{1}{\sqrt{K^2}}(K^0,0,0,|\vec{K}|)\nonumber\\
 \kappa_{Q}^\mu(1) &=&(0,1,0,0) \nonumber\\
  \kappa_{Q}^\mu(2) &=& (0,0,1,0)\nonumber\\
  \kappa_{Q}^\mu(0) &=&\frac{1}{\sqrt{K^2}}(|\vec{K}|,0,0,K^0)\,,
\label{eq:K2Q}
\end{eqnarray}
Similarly, the polarization vectors of the $Y$ system in the $Q$ rest frame,
with $\epsilon_{Q}^\mu(m)$ for $m=1,2,0$, can be written as
\begin{eqnarray}
  \epsilon_{Q}^\mu(t) &=& \frac{1}{\sqrt{Y^2}}(Y^0,0,0,-|\vec{K}|)\nonumber\\
 \epsilon_{Q}^\mu(1) &=&(0,-1,0,0) \nonumber\\
  \epsilon_{Q}^\mu(2) &=& (0,0,-1,0)\nonumber\\
  \epsilon_{Q}^\mu(0) &=&\frac{1}{\sqrt{Y^2}}(|\vec{K}|,0,0,-Y^0)\,.
\label{eq:R2Q}
\end{eqnarray}
In the rest frame of $Y$, we have
\begin{eqnarray}
  Y^\mu &=&(\sqrt{Y^2},0,0,0)\,, \nonumber\\
  q_{12}^\mu &=& (q_1-q_2)^\mu =(E_1-E_2,2|\vec{q}_{1}|\sin\theta_{1}\cos\phi_{1},-2|\vec{q}_{1}|\sin\theta_{1}\sin\phi_{1},-2|\vec{q}_{1}|\cos\theta_{1})\,,
\end{eqnarray}
and the polarization vectors for the $Y$ system in its own rest frame read
\begin{eqnarray}
  \epsilon_{Y}^\mu(t) &=& (1,0,0,0)\,,\nonumber\\
 \epsilon_{Y}^\mu(1) &=&(0,-1,0,0)\,,\nonumber\\
  \epsilon_{Y}^\mu(2) &=& (0,0,-1,0)\,,\nonumber\\
  \epsilon_{Y}^\mu(0) &=&(0,0,0,-1)\,,
\label{eq:R2R}
\end{eqnarray}
where the minus sign is chosen to align the momentum and polarization
direction. In the rest frame of $K$, the relevant four-momenta are
\begin{eqnarray}
  K^\mu &=&(\sqrt{K^2},0,0,0)\,, \nonumber\\
  Z^\mu &=&(Z^0,|\vec{Z}|\sin\theta_{2},0,|\vec{Z}|\cos\theta_{2})\,.
\end{eqnarray}
The polarization vectors for the $K$ system in its own rest frame is
\begin{eqnarray}
  \kappa_{K}^\mu(t) &=&(1,0,0,0)\nonumber\\
 \kappa_{K}^\mu(1) &=&(0,1,0,0) \nonumber\\
  \kappa_{K}^\mu(2) &=& (0,0,1,0)\nonumber\\
  \kappa_{K}^\mu(0) &=&(0,0,0,1)\,,
\label{eq:K2K}
\end{eqnarray}
and these for the $Z$ system in the $K$ rest frame can be given by
\begin{eqnarray}
  \eta_{K}^\mu(t) &=&\frac{1}{\sqrt{Z^2}}( Z^0,|\vec{Z}|\sin\theta_{2},0,|\vec{Z}|\cos\theta_{2})\,,\nonumber\\
 \eta_{K}^\mu(1) &=&(0,\cos\theta_{2},0,-\sin\theta_{2})\,,\nonumber\\
  \eta_{K}^\mu(2) &=& (0,0,1,0)\,,\nonumber\\
  \eta_{K}^\mu(0) &=&\frac{1}{\sqrt{Z^2}}(|\vec{Z}|, Z^0\sin\theta_{2},0,Z^0\cos\theta_{2})\,,
\label{eq:N2K}
\end{eqnarray}
where the condition $\eta_{K}^\mu(m)Z_\mu$ for $m=1,2,0$ is required. Finally,
in the rest frame of $Z$, we have
\begin{eqnarray}
  Z^\mu &=&(\sqrt{Z^2},0,0,0) \nonumber\\
  q_{45}^\mu &=&( E_4-E_5,2|\vec{q}_{4}|(\cos\theta_{2}\sin\theta_{3}\cos\phi_{3}+\sin\theta_{2}\cos\theta_{3})\,,\nonumber\\
  &&2|\vec{q}_{4}|\sin\theta_{3}\sin\phi_{3},2|\vec{q}_{4}|(\cos\theta_{2}\cos\theta_{3}-\sin\theta_{2}\sin\theta_{3}\cos\phi_{3}))\,,
\end{eqnarray}
and the polarization vectors for the $Z$ system in its own rest frame can be
written as
\begin{eqnarray}
  \eta_{Z}^\mu(t) &=&(1,0,0,0)\,,\nonumber\\
 \eta_{Z}^\mu(1) &=&(0,\cos\theta_{2},0,-\sin\theta_{2})\,,\nonumber\\
  \eta_{Z}^\mu(2) &=& (0,0,1,0)\,,\nonumber\\
  \eta_{Z}^\mu(0) &=&(0, \sin\theta_{2},0,\cos\theta_{2})\,.
\label{eq:N2N}
\end{eqnarray}
To clarify, it is important to note that each set of polarization vectors,
namely
$\{\epsilon_Q,\epsilon_Y\}$, $\{\eta_Z,\eta_K\}$, and
$\{\kappa_K,\kappa_Q\}$, satisfies the orthonormality and completeness
properties as described in Eq.~(\ref{eq:orthonormality}).

\section{Decay rate formalism}
\begin{figure}[t!]
  \centering
  \includegraphics[width=4.0in]{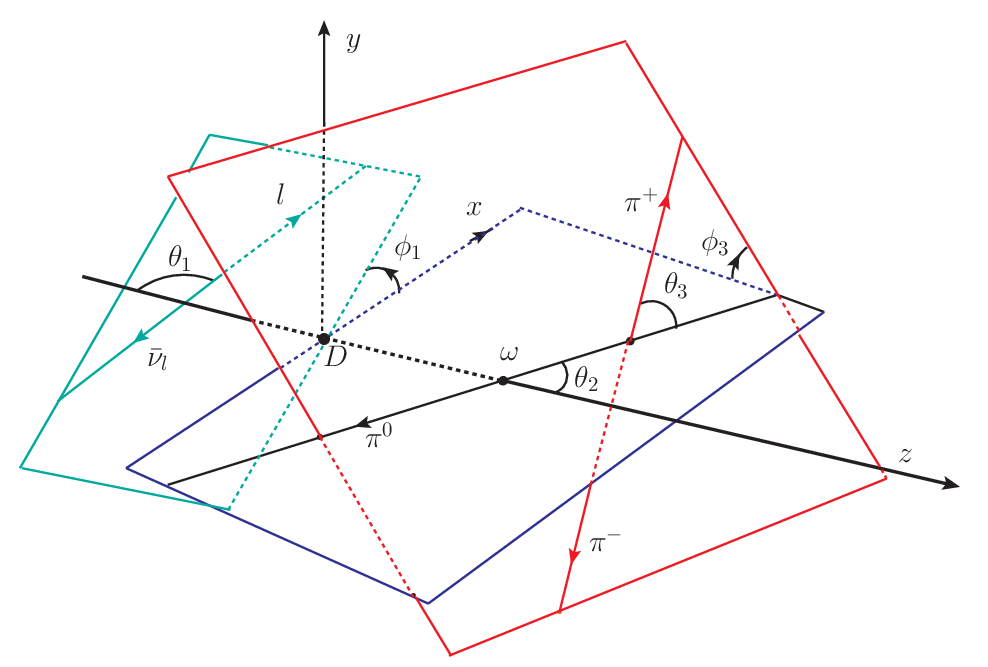}
  \caption{Definition of angles in the cascade decay
    $D\to V\ell^+\nu_{\ell}, V\to P_1P_2P_3$. $\theta_1$~($\theta_2$) is the
    angle between the three-momenta of $\ell^+$~($P_1$) and $D$ in the
    $\ell^+\nu_{\ell}$~($V$) rest frame; $\theta_3$ is the angle between the
    three-momenta of $P_3$ and $V$ in the $P_2P_3$ rest frame;
    $\phi_1$~($\phi_3$) is the angle between the $\ell\nu_{\ell}$~($P_2P_3$)
    and the $VP_1$ systems.}
  \label{fig:angle}
\end{figure}

The decay rate of a $D\to V\ell^+\nu_{\ell}, V\to P_1P_2P_3$
decay\footnote{We utilize the decay process of the $D$ meson as a specific
example to showcase the derivation of the pertinent equations. The calculation
methodology remains identical for the $D_s$, $B$, and $B_s$ mesons, with the
only difference being the substitution of the corresponding physical
parameters.} can be parameterized using the following eight independent
variables as illustrated in Fig.~\ref{fig:angle}:
\begin{itemize}
\item $Y^2$: the mass squared of the $\ell^+\nu_{\ell}$ system
\item $K^2$: the mass squared of the $P_1P_2P_3$ system
\item $Z^2$: the mass squared of the $P_2P_3$ system
\item $\theta_1$: the angle between the three-momenta of $\ell^+$ in the
  $\ell^+\nu_{\ell}$ rest frame and the line of the flight of the
  $\ell^+\nu_{\ell}$ system in the $D$ rest frame
\item $\theta_2$: the angle between the three-momenta of $P_1$ in the $V$ rest
  frame and the line of the flight of $V(P_1P_2P_3)$ in the $D$ rest frame
\item $\theta_3$: the angle between the three-momenta of $P_3$ in the $P_2P_3$
  rest frame and the line of the flight of the $P_2P_3$ system in the
  $V(P_1P_2P_3)$ rest frame
\item $\phi_1$: the angle between the planes formed in the $D$ rest frame by
  the momenta of the $\ell^+\nu_{\ell}$ pair and the $V(P_1P_2P_3)P_1$ pair.
  The sense of $\phi_1$ is from the $V(P_1P_2P_3)P_1$ plane to the
  $\ell^+\nu_{\ell}$ plane.
\item $\phi_3$: the angle between the planes formed in the $V(P_1P_2P_3)$ rest
  frame by the momenta of the $P_2P_3$ pair and $DP_1$ pair, respectively. The
  sense of $\phi_3$ is from the $DP_1$ plane to the $P_2P_3$ plane.
\end{itemize}
The four-momenta of $\ell^+$, $\nu_{\ell}$, $P_1$, $P_2$, $P_3$ are denotes as
$q_{\ell}$, $q_\nu$, $p_1$, $p_2$, and $p_3$ (corresponding to $q_{1\cdots5}$
in Eq.~(\ref{eq:Gamma_5}), and their masses as $m_{\ell}$, $m_\nu$, $m_1$,
$m_2$, and $m_3$, respectively. For convenience, we define
$Q\equiv q_{\ell}+q_\nu+p_1+p_2+p_3$, $Y\equiv q_{\ell}+q_\nu$,
$K\equiv p_1+p_2+p_3$, and $Z\equiv p_2+p_3$, and denote the masses of $D$ and
$V$ as $m_A$ and $m_V$, respectively.

\subsection{Amplitude}
The decay amplitude
${\cal A}_p\equiv{\cal A}(D\to V\ell^+\nu_{\ell}, V\to P_1P_2P_3)$ can be
decomposed into two parts: $V\to P_1P_2P_3$ and $D\to V\ell^+\nu_{\ell}$. We
first focus on the amplitude of $V\to P_1P_2P_3$:
\begin{eqnarray}
  {\cal A}(V\to P_1P_2P_3) &=& g\epsilon_{\mu\nu\alpha\beta}\epsilon_V^{\mu}p_1^\nu p_2^\alpha p_3^\beta=-\frac{1}{2}g\epsilon_{\mu\nu\alpha\beta}\epsilon_V^{\mu}K^\nu Z^\alpha p_{23}^\beta\,,
\label{ea:A_V2PPP}
\end{eqnarray}
with $p_{23}=p_{2}-p_{3}$. The coupling incorporates information about the
decay structure of $V$ can always be expressed in terms of $Z^2$ and $\theta_3$,
$g=g(Z^2, \theta_3)$. For the specific case of $\omega \to \pi^+\pi^-\pi^0$,
\begin{eqnarray}
  g =g_{3\pi}+g_{\rho\pi}g_{\rho\to\pi\pi}(D^{-1}[\rho^0, p_{\pi^+}+p_{\pi^-}]+D^{-1}[\rho^+,p_{\pi^0}+p_{\pi^+}]+D^{-1}[\rho^-,p_{\pi^0}+p_{\pi^-}])\,,
\end{eqnarray}
where $D[\rho,p]=p^2-m_\rho^2+im_\rho\Gamma_\rho$. It is known that
$(p_{\pi^+}+p_{\pi^-})^{2}$, $(p_{\pi^0}+p_{\pi^+})^{2}$ and
$(p_{\pi^0}+p_{\pi^-})^{2}$ can be expressed as functions of $Z^2$ and
$\theta_3$~\cite{Gudino:2011ri}.

Equation~(\ref{ea:A_V2PPP}) involves two frames: the $Z$-frame conveniently
expresses $p_{23}$, whereas $\epsilon_V$, $K$, and $Z$ can be readily expressed
in the $K$-frame. By utilizing the orthonormality and completeness relations of
polarization vectors~(Eq.~(\ref{eq:orthonormality})), it is able to establish a
connection between the two frames through the introduction of the following
definitions:
\begin{eqnarray}
  {\cal C}^{\lambda_m}&=& \epsilon_{\mu\nu\alpha\beta}\epsilon_V^{\mu}(\lambda_m)K^\nu Z^\alpha p_{23}^\beta ={\cal T}^{\lambda_m}_{\lambda_n}{\cal S}_{\lambda_n}g_{nn}\,,
\end{eqnarray}
where
\begin{eqnarray}
  {\cal T}^{\lambda_m}_{\lambda_n}&=&\epsilon_{\mu\nu\alpha\beta}\epsilon_V^{\mu}(\lambda_m)K^\nu Z^\alpha\eta{K}^{\dagger\beta}(\lambda_n)\,,\nonumber\\
  {\cal S}_{\lambda_n}&=&\eta^{\mu^{\prime}}_{Z}(\lambda_n) p_{23,{\mu^{\prime}}}\,.
\end{eqnarray}
With Eqs.~(\ref{eq:N2K})-(\ref{eq:N2N}), the non-vanishing terms of
${\cal T}^{\lambda_m}_{\lambda_n}$ and ${\cal S}_{\lambda_n}$ are
\begin{eqnarray}
  {\cal T}^{0}_{\pm} &=&\frac{ i}{\sqrt{2}}|\vec{Z}|\sqrt{K^2}\sin\theta_2\,,\nonumber\\
  {\cal T}^{+}_{\pm} &=&\frac{ i}{2}|\vec{Z}|\sqrt{K^2}(\cos\theta_2\pm1)\,,\nonumber\\
  {\cal T}^{-}_{\pm} &=&-\frac{ i}{2}|\vec{Z}|\sqrt{K^2}(\cos\theta_2\mp1)\,,\nonumber\\
  {\cal S}_{\pm} &=& \pm\sqrt{2}|\vec{p}_{2}|\sin\theta_3e^{\pm i\phi_3}\,.\label{ea:c_V2PPP}
\end{eqnarray}
As a result, we obtain
\begin{eqnarray}
  {\cal C}^{0} &=& -{\cal T}^{0}_{\pm}{\cal S}_{\pm}=2|\vec{Z}|\sqrt{K^2}|\vec{p}_{2}|\sin\theta_2\sin\theta_3\sin\phi_3\,,\nonumber\\
  {\cal C}^{+} &=& -{\cal T}^{+}_{\pm}{\cal S}_{\pm}=-i\sqrt{2}|\vec{Z}|\sqrt{K^2}|\vec{p}_{2}|(\cos\phi_3+ i\cos\theta_2\sin\phi_3)\sin\theta_3\,,\nonumber\\
  {\cal C}^{-} &=& -{\cal T}^{-}_{\pm}{\cal S}_{\pm}=-i\sqrt{2}|\vec{Z}|\sqrt{K^2}|\vec{p}_{2}|(\cos\phi_3- i\cos\theta_2\sin\phi_3)\sin\theta_3\,,
\end{eqnarray}

Next, we consider the amplitude for semileptonic decay of $D$ to $V$, which is
given by
\begin{eqnarray}
  {\cal A}(D\to V\ell^+\nu_{\ell}) &=& \frac{G_F}{\sqrt{2}}V_{cq}\langle V|\bar{q}\gamma_{\mu}(1-\gamma_{5})c|D\rangle \bar{u}(q_{\ell})\gamma^{\mu}(1-\gamma_{5})v(q_\nu)\,,\nonumber\\
  &=&\frac{G_F}{\sqrt{2}}V_{cq}({\cal H}_V^\mu-{\cal H}_A^\mu) {\cal L}_\mu\,.
\end{eqnarray}
Here, ${\cal H}_V^\mu=\langle V|\bar{q}\gamma^\mu c|D\rangle$,
${\cal H}_A^\mu=\langle V|\bar{q}\gamma^\mu\gamma_5c|D\rangle$,
$G_F=1.166\times10^{-5}$~GeV$^{-2}$ represents the
Fermi constant, and $V_{cq}$ denotes the Cabibbo-Kobayashi-Maskawa matrix
element. Inserting polarization vectors to connect the $Y$- and
$Q$-frames~(Eqs.~(\ref{eq:R2Q})-(\ref{eq:R2R})), the leptonic part is written
as
\begin{eqnarray}
  {\cal L}^{\lambda_l}_{\lambda_W} &=&\epsilon_{Y,\mu}(\lambda_W)\bar{u}_{\lambda_l}(q_{\ell})\gamma^{\mu}(1-\gamma_{5})v(q_\nu)\,,
\end{eqnarray}
with the non-vanishing terms~\cite{Mandal:2020htr, Becirevic:2019tpx}
\begin{eqnarray}
  {\cal L}^{+}_{+} &=&-\sqrt{2}m_{\ell}\beta_{\ell}\sin\theta_{1}e^{-2i\phi_1}\,,\nonumber\\
  {\cal L}^{+}_{-} &=&\sqrt{2}m_{\ell}\beta_{\ell}\sin\theta_{1}\,,\nonumber\\
  {\cal L}^{+}_{0} &=&2m_{\ell}\beta_{\ell}\cos\theta_{1}e^{-i\phi_1}\,,\nonumber\\
  {\cal L}^{+}_{t} &=&-2m_{\ell}\beta_{\ell}e^{-i\phi_1}\,,\nonumber\\
  {\cal L}^{-}_{\pm} &=&\sqrt{2Y^2}\beta_{\ell}(1\pm\cos\theta_{1})e^{\mp i\phi_1}\,,\nonumber\\
  {\cal L}^{-}_{0} &=&2\sqrt{Y^2}\beta_{\ell}\sin\theta_{1}\,,
\end{eqnarray}
with $\beta_{\ell}=\sqrt{1-\frac{m_{\ell}^2}{Y^2}}$. The hadronic part is
written as
\begin{eqnarray}
  {\cal H}^{\lambda_m}_{V(A),\lambda_W} &=&\epsilon^\dagger_{Q,\mu}(\lambda_W)\langle V(\lambda_m)\mid V^{\mu}(A^{\mu})\mid D\rangle\,,
\end{eqnarray}
with the non-vanishing terms
\begin{eqnarray}
  {\cal H}^{\pm}_{V,\pm} &=&\mp2\frac{m_A}{m_A+m_V}|\vec{K}|V_0\,,\nonumber\\
  {\cal H}^{\pm}_{A,\pm} &=&(m_A+m_V)A_1\,,\nonumber\\
  {\cal H}^{0}_{A,0} &=&-8\frac{m_Am_V}{\sqrt{Y^2}}A_{12}\,,\nonumber\\
  {\cal H}^{0}_{A,t} &=&-2\frac{m_A}{\sqrt{Y^2}}|\vec{K}|A_0\,,
  \label{eq:HVA}
\end{eqnarray}
where
\begin{eqnarray}
  A_{12} &=& \frac{1}{16m_Am^2_V} (m^2_A-m^2_V-Y^2)(m_A+m_V)A_1-4\frac{m^2_A}{m_A+m_V}|\vec{K}|^2 A_2\,.
\end{eqnarray}
Here, $V_0$ and $A_{0,1,2}$ are the form factors of the vector meson. Further
discussion regarding the form factors can be found in Section~\ref{sec:num}.

\subsection{Decay rate}
Combining the element of five-body phase space in Eq.~(\ref{eq:Gamma_5}) and
collecting amplitudes in Eqs.~(\ref{ea:A_V2PPP})-(\ref{eq:HVA}), one have the
decay rate of $D\to V\ell^+\nu_{\ell}, V\to P_1P_2P_3$ as
\begin{eqnarray}
  \Gamma_{5} &=& \frac{1}{2^{17} \pi^{9}m_A^2}\int\frac{|\vec{q}_{1}||\vec{q}_{4}||\vec{Z}||\vec{K}|}{\sqrt{K^2}\sqrt{Y^2}\sqrt{Z^2}}\sum_{\lambda_l}|{\cal A}_p^{\lambda_l}|^2d\Omega_{1}d\Omega_{3}d\cos\theta_{2} d K^2 d Y^2 d Z^2\,,
\end{eqnarray}
with
\begin{eqnarray}
  {\cal A}_p^{\lambda_l} &=&-\frac{G_F}{\sqrt{2}}V_{q_1q_2}\sum_{\lambda_1}g_ {\lambda_1\lambda_1}{\cal L}^{\lambda_l}_{V-A,\lambda_1}\left[\sum_{\lambda_2}\frac{g}{2}\frac{({\cal H}^{\lambda_2}_{V,\lambda_1}+{\cal H}^{\lambda_2}_{A,\lambda_1})}{K^2-m^2_V+im_V\Gamma_V} {\cal C}^{\lambda_2}\right]\,.
\end{eqnarray}
Consequently, ${\cal A}_p^{\lambda_l}$ can be expressed in terms of the form
factors of the vector meson:
\begin{eqnarray}
  {\cal A}_p^{+} &=&-\frac{g}{\sqrt{2}}|\vec{Z}||\vec{q}_{4}||\sqrt{K^2}|\sin\theta_3\frac{G_F}{\sqrt{2}}V_{cd}(F_1\alpha_1+F_2\alpha_2+F_3\alpha_3+F_4\alpha_4)\,,\nonumber\\
  {\cal A}_p^{-} &=&-\frac{g}{\sqrt{2}}|\vec{Z}||\vec{q}_{4}||\sqrt{K^2}|\sin\theta_3\frac{G_F}{\sqrt{2}}V_{cd}(F^\prime_1\beta_1+F^\prime_2\beta_2+F^\prime_3\beta_3)\,,
\end{eqnarray}
where
\begin{eqnarray}
  F_1&=& 2\sqrt{2}m_{\ell}\beta_{\ell}\frac{m_A}{m_A+m_V}V_0|\vec{K}|\,,\nonumber\\
  F_2&=& -\sqrt{2}m_{\ell}\beta_{\ell}(m_A+m_V)A_1\,,\nonumber\\
  F_3 &=& 16\sqrt{2}m_{\ell}\beta_{\ell}\frac{m_A m_V}{\sqrt{Y^2}}A_{12}\,,\nonumber\\
  F_4 &=& 4\sqrt{2}m_{\ell}\beta_{\ell}\frac{m_A }{\sqrt{Y^2}}|\vec{K}|A_{0}\,,\nonumber\\
  F^\prime_1 &=& -2\sqrt{2}\sqrt{Y^2}\beta_{\ell}\frac{m_A}{m_A+m_V}V_0|\vec{K}|\,,\nonumber\\
  F^\prime_2 &=&\sqrt{2}\sqrt{Y^2}\beta_{\ell}(m_A+m_V)A_1\,,\nonumber\\
  F^\prime_3 &=& 16\sqrt{2}\beta_{\ell}m_A m_V A_{12}
\end{eqnarray}
and
\begin{eqnarray}
  \alpha_1&=& i\sin\theta_1e^{-2i\phi_1}(\cos\phi_3+i\cos\theta_2\sin\phi_3)+i\sin\theta_1(\cos\phi_3-i\cos\theta_2\sin\phi_3)\,,\nonumber \\
  \alpha_2&=&i\sin\theta_1e^{-2i\phi_1}(\cos\phi_3+i\cos\theta_2\sin\phi_3)-i\sin\theta_1(\cos\phi_3-i\cos\theta_2\sin\phi_3)\,,\nonumber\\
  \alpha_3 &=& \cos\theta_1 e^{-i\phi_1}\sin\theta_2\sin\phi_3\,,\nonumber\\
  \alpha_4 &=& e^{-i\phi_1}\sin\theta_2\sin\phi_3\,,\nonumber\\
  \beta_1 &=& i(1+\cos\theta_1)e^{-i\phi_1}(\cos\phi_3+i\cos\theta_2\sin\phi_3)-i(1-\cos\theta_1)(\cos\phi_3-i\cos\theta_2\sin\phi_3)\,,\nonumber\\
  \beta_2 &=&i(1+\cos\theta_1)e^{i\phi_1}(\cos\phi_3+i\cos\theta_2\sin\phi_3)+i(1-\cos\theta_1)(\cos\phi_3-i\cos\theta_2\sin\phi_3)\,,\nonumber\\
  \beta_3 &=& \sin\theta_1\sin\theta_2\sin\phi_3\,.
\end{eqnarray}

Next, we derive the decay rate in the narrow width limit
$\Gamma_V\rightarrow 0$. In this case, one can use approximation
$\frac{1}{|K^2-m^2_V+i m_V\Gamma_V|^2}\rightarrow \frac{\pi}{m_V\Gamma_V}\delta(K^2-m^2_V)$
and integrate over the mass dependence, i.e. $K^2$, in the phase space. The
decay rate is then given by
\begin{eqnarray}\label{eq:omega_width}
  \Gamma_{5}
  &=& \frac{3G^2_F|V_{q_1q_2}|^2}{2^{12} \pi^{5}m_A^2\Gamma_V}\int\frac{4g^2}{3}|\vec{q}_{4}|^3|\vec{Z}|^3\frac{\sin^2\theta_3}{(8\pi)^3\sqrt{Z^2}}\frac{|\vec{q}_{1}||\vec{K}|}{\sqrt{Y^2}}(I+I^{\prime}) d\Omega_{1}d\Omega_{3}d\cos\theta_{2} d Y^2d Z^2\,.
\end{eqnarray}
Since the specific form of the effective coupling constant $g$ remains unknown,
the equation above is merely a formal expression. Fortunately, $g$ is dependent
on $Z^2$ and $\theta_3$. By integrating over $Z^2$ and $\theta_3$, the terms
involving $Z^2$ and $\theta_3$ can be substituted with the branching fraction of
$V\to P_1P_2P_3$, denoted as ${\cal}{Br}$, which is written
as~\cite{Faessler:2002ut}
\begin{eqnarray}
  {\cal}{Br}(V\to P_1P_2P_3) &=& \frac{{\cal }{\Gamma}(V\to P_1P_2P_3)}{{\cal }{\Gamma}_V}
  =\frac{1}{{\cal }{\Gamma}_V}\int\frac{4m_V|\vec{q}_{2}||\vec{Z}|}{(8\pi m_V)^3\sqrt{Z^2}}\overline{\mid{\cal A}\mid^2}d\cos\theta_{3}dZ^2\,.\label{eq:A1}
\end{eqnarray}
With the help of Eqs.~(\ref{ea:A_V2PPP})-(\ref{ea:c_V2PPP}), one can obtain the
average value of the decay amplitude squared 
\begin{eqnarray}
  \overline{\mid{\cal A}\mid^2} &=&\frac{g^2}{3} |\vec{q}_{2}|^2|\vec{Z}|^2 K^2 \sin^2\theta_3\,,\label{eq:A2}
\end{eqnarray}
and the branching fraction of $V\to P_1P_2P_3$
\begin{eqnarray}
  {\cal}{Br}(V\to P_1P_2P_3) &=& \frac{1}{{\cal}{\Gamma_V}}
    \int\frac{4g^2|\vec{q}_{2}|^3|\vec{K}|^3}{3(8\pi )^3\sqrt{Z^2}}\sin^2\theta_3 d\cos\theta_{3}dZ^2\,.\label{eq:A3}
\end{eqnarray}
Substituting Eq.~(\ref{eq:A3}) into Eq.~(\ref{eq:omega_width}), one can obtain
\begin{eqnarray}\label{eq:omega_width2}
  \Gamma_{5}
  &=& \frac{3G^2_F|V_{q_1q_2}|^2{\cal}{Br}}{2^{12} \pi^{5}m_A^2}\int\frac{|\vec{q}_{1}||\vec{K}|}{\sqrt{Y^2}}(I+I^{\prime}) d\Omega_{1} d\phi_{3} d\cos\theta_{2} dY^2\,.
\end{eqnarray}
The presence of five degrees of freedom in Eq.~\ref{eq:omega_width2} can be
comprehended from another perspective that involves polarization information.
Firstly, a simple three-body decay has two degrees of freedom. Then, $\omega$
has three polarizations, and the branching fraction only restricts the sum of
the magnitudes of the three polarization modes, leaving two degrees of freedom
in terms of polarization magnitude. Moreover, the longitudinal polarization
direction is always aligned with the momentum direction of $\omega$, which has
no degrees of freedom, while the transverse polarization direction can rotate
within the plane perpendicular to the longitudinal polarization, adding one
degree of freedom.

The decay intensity $I^{(\prime)}$ can be decomposed with respect to $\theta_1$
and $\phi_1$, written as
\begin{eqnarray}
  I^{(\prime)} &=& I^{(\prime)}_1+I^{(\prime)}_2\cos2\theta_1+I^{(\prime)}_3\sin^2\theta_1\cos2\phi_1+I^{(\prime)}_4\sin2\theta_1\cos\phi_1+I^{(\prime)}_5\sin\theta_1\cos\phi_1+I^{(\prime)}_6\cos\theta_1\nonumber\\
  &&+I^{(\prime)}_7\sin\theta_1\sin\phi_1+I^{(\prime)}_8\sin2\theta_1\sin\phi_1+I^{(\prime)}_9\sin^2\theta_1\sin2\phi_1\,,
\end{eqnarray}
where $I^{(\prime)}_{1,...,9}$ can be expressed in terms of the form factors
$F_1^{(\prime)}$, $F_2^{(\prime)}$, $F_3^{(\prime)}$:
\begin{eqnarray}
  I_1 &=& (|F_1|^2 +|F_2|^2)(\cos^2\phi_3+\cos^2\theta_2\sin^2\phi_3)+(\frac{1}{2}|F_3|^2+|F_4|^2)\sin^2\theta_2\sin^2\phi_3\,,\nonumber\\
  I_2 &=& -(|F_1|^2 +|F_2|^2)(\cos^2\phi_3+\cos^2\theta_2\sin^2\phi_3)+\frac{1}{2}|F_3|^2\sin^2\theta_2\sin^2\phi_3\,,\nonumber\\
  I_3 &=&2(|F_1|^2 -|F_2|^2)(\cos^2\phi_3-\cos^2\theta_2\sin^2\phi_3)\,,\nonumber\\
  I_4 &=&-F_2F_3\sin2\theta_2 \sin^2\phi_3\,,\nonumber\\
  I_5 &=& -2F_2F_4\sin2\theta_2 \sin^2\phi_3\,,\nonumber\\
  I_6 &=&2F_3F_4\sin^2\theta_2 \sin^2\phi_3\,,\nonumber\\
  I_7&=& 2F_2F_4\sin\theta_2 \sin2\phi_3\,,\nonumber\\
  I_8 &=&F_2F_3\sin\theta_2 \sin2\phi_3\,,\nonumber\\
  I_9 &=&2(|F_1|^2 -|F_2|^2)\cos\theta_2 \sin2\phi_3\,,\nonumber\\
  I^{\prime}_1 &=& 3(|F^{\prime}_1|^2 +|F^{\prime}_2|^2)(\cos^2\phi_3+\cos^2\theta_2\sin^2\phi_3)+\frac{1}{2}|F^{\prime}_3|^2\sin^2\theta_2\sin^2\phi_3\,,\nonumber\\
  I^{\prime}_2 &=& (|F^{\prime}_1|^2 +|F^{\prime}_2|^2)(\cos^2\phi_3+\cos^2\theta_2\sin^2\phi_3)-\frac{1}{2}|F^{\prime}_3|^2\sin^2\theta_2\sin^2\phi_3\,,\nonumber\\
  I^{\prime}_3 &=&-2(|F^{\prime}_1|^2 -|F^{\prime}_2|^2)(\cos^2\phi_3-\cos^2\theta_2\sin^2\phi_3)\,,\nonumber\\
  I^{\prime}_4 &=&-F^{\prime}_2F^{\prime}_3\sin2\theta_2 \sin^2\phi_3\,,\nonumber\\
  I^{\prime}_5 &=& -2F^{\prime}_1F^{\prime}_3\sin2\theta_2 \sin^2\phi_3\,,\nonumber\\
  I^{\prime}_6 &=&8F^{\prime}_1F^{\prime}_2(\cos^2\phi_3+\cos^2\theta_2\sin^2\phi_3)\,,\nonumber\\
  I^{\prime}_7&=& -2F^{\prime}_1F^{\prime}_3\sin\theta_2 \sin2\phi_3\,,\nonumber\\
  I^{\prime}_8 &=&-F^{\prime}_2F^{\prime}_3\sin\theta_2 \sin2\phi_3\,,\nonumber\\
  I^{\prime}_9 &=&2(|F^{\prime}_1|^2 -|F^{\prime}_2|^2)\cos\theta_2 \sin2\phi_3\,.
\end{eqnarray}

Additionally, as a cross-verification by reproducing the standard three-body
formalism, one can derive the angular distribution for the three-body
$D\to V\ell^+\nu_\ell$ decay by integrating over the phase space related to the
$V$ system:
\begin{eqnarray}
  \frac{ d\Gamma_{5}}{dY^2d\cos\theta_{1}} &=&{\cal}{Br}\frac{ d\Gamma(D\to V\ell^+\nu_\ell)}{dY^2d\cos\theta_{1}}\,.
\end{eqnarray}
with
\begin{eqnarray}
  \frac{ d\Gamma(D\to V\ell^+\nu_\ell)}{dY^2d\cos\theta_{1}} &=& \frac{3G^2_F|V_{q_1q_2}|^2}{2^{12} \pi^{5}m_A^2}\int\frac{|\vec{q}_{1}||\vec{K}|}{\sqrt{Y^2}}(I+I^{\prime}) d\phi_{1} d\phi_{3} d\cos\theta_{2}\nonumber\\
 &=& \frac{3G^2_F|V_{q_1q_2}|^2{\cal}{Br}}{2^{11} \pi^{4}m_A^2}\frac{|\vec{q}_{1}||\vec{K}|}{\sqrt{Y^2}}\bigg\{\left[\left(|F_1|^2 +|F_2|^2\right)\frac{8\pi}{3}+\left(\frac{1}{2}|F_3|^2+|F_4|^2\right)\frac{4\pi}{3}\right]\nonumber\\
 &+&\left[-\left(|F_1|^2 +|F_2|^2\right)\frac{8\pi}{3}+\left(\frac{1}{2}|F_3|^2+|F_4|^2\right)\frac{4\pi}{3}\right]\cos2\theta_1+F_3F_4\frac{8\pi}{3}\cos\theta_1\nonumber\\
 &+&\left[3\left(|F^{\prime}_1|^2 +|F^{\prime}_2|^2\right)\frac{8\pi}{3}+\frac{1}{2}|F^{\prime}_3|^2\frac{4\pi}{3}\right]\nonumber\\
 &+&\left[\left(|F^{\prime}_1|^2 +|F^{\prime}_2|^2\right)\frac{8\pi}{3}-\frac{1}{2}|F^{\prime}_3|^2\frac{4\pi}{3}\right]\cos2\theta_1+8F^{\prime}_1F^{\prime}_2\frac{8\pi}{3}\cos\theta_1\bigg\}\,,
\end{eqnarray}
which is equivalent to the results in Refs.~\cite{Fajfer:2005ug,Celis:2012dk}.

\section{NUMERICAL RESULTS AND DISCUSSIONS}
\label{sec:num}
The dependence on a variable is lost after integration over the variable, but
it can be recovered by considering the asymmetry.
Reference~\cite{Mandal:2020htr} has discussed several asymmetries, including
the well-known forward-backward asymmetry, angular asymmetries, etc. With the
extra degrees of freedom, five-body decays exhibit unique asymmetries that can
be used to test different parametrizations of form factors. The following
correlations are considered:
\begin{itemize}
\item the distributions of $\cos\theta_1$ and $\cos\theta_2$ or
  \begin{eqnarray}
    \int_{D_{1}}d\cos\theta&\equiv&\left(\int_{-1}^0+\int_0^1\right)d\cos\theta\,,\nonumber\\
    \int_{D_{2}}d\cos\theta&\equiv&\left(\int_{-1}^0-\int_0^1\right)d\cos\theta\,,
  \end{eqnarray}
\item the distributions of $\phi_{1}$ and $\phi_{3}$
  \begin{eqnarray}
    \int_{S_{1}}d\phi&\equiv&\left(\int_{0}^\pi+\int_\pi^{2\pi}\right)d\phi\,,\nonumber\\
    \int_{S_{2}}d\phi&\equiv&\left(\int_{0}^\pi-\int_\pi^{2\pi}\right)d\phi\,,\nonumber\\
    \int_{S_{3}}d\phi&\equiv&\left(\int_{0}^{\pi/2}-\int_{\pi/2}^\pi+\int_\pi^{3\pi/2}-\int_{3\pi/2}^{2\pi}\right)d\phi\,.
  \end{eqnarray}
\end{itemize}
These correlations relate to the asymmetry of corresponding variables. We
investigate the following three parameters, which are non-vanishing in the
standard model:
\begin{eqnarray}
  \Upsilon_1 &=&\frac{\int_{S_{3}}d\phi_{1}\int_{S_{2}}d\phi_{3} \int_{D_1} d\cos\theta_{2}\int_{D_1} d\cos\theta_{1}d\Phi_5}{\int_{S_1} d\phi_{1}\int_{S_1}d\phi_{3} \int_{D_1} d\cos\theta_{2}\int_{D_1} d\cos\theta_{1}d\Phi_5}\,,\nonumber\\
  \Upsilon_2 &=& \frac{\int_{S_{3}}d\phi_{1}\int_{S_{3}}d\phi_{3} \int_{D_2} d\cos\theta_{2}\int_{D_1} d\cos\theta_{1}d\Phi_5}{\int_{S_1} d\phi_{1}\int_{S_1}d\phi_{3} \int_{D_1} d\cos\theta_{2}\int_{D_1} d\cos\theta_{1}d\Phi_5}\,,\nonumber\\
  \Upsilon_3 &=& \frac{\int_{S_{3}}d\phi_{1}\int_{S_{2}}d\phi_{3} \int_{D_1} d\cos\theta_{2}\int_{D_2} d\cos\theta_{1}d\Phi_5}{\int_{S_1} d\phi_{1}\int_{S_1}d\phi_{3} \int_{D_1} d\cos\theta_{2}\int_{D_1} d\cos\theta_{1}d\Phi_5}\,,
\end{eqnarray}
where
$d\Phi_5\equiv\frac{d\Gamma_5}{d\Omega_{1} d\phi_{3} d\cos\theta_{2} dY^2}$.

The effective three-body semileptonic phase space can't be used to investigate
the coherent terms of polarizations, but, with the additional angular
distributions, the five-body phase space can. Take $\int_{S_{3}}d\phi$ as an
example, one has
\begin{eqnarray}
  d\Phi_5\equiv\frac{d\Gamma_5}{d\Omega_{1} d\phi_{3} d\cos\theta_{2} dY^2}\propto \mid{\sum_{i=0,\pm}{\cal A}(D\to V(i)\ell^+\nu_{\ell}){\cal A}(V(i)\to P_1P_2P_3)}\mid^2\,,
\end{eqnarray}
and
\begin{eqnarray}\label{eq:coherent}
  &&\mid{\sum_{i=0,\pm}{\cal A}(D\to V(i)\ell^+\nu_{\ell}){\cal A}(V(i)\to P_1P_2P_3)}\mid^2 \,,\nonumber\\
  &=&\sum_{i=0,\pm}\mid{{\cal A}(D\to V(i)\ell^+\nu_{\ell}){\cal A}(V(i)\to P_1P_2P_3)}\mid^2 \,,\\
  &+&\sum_{i\neq j}{\cal A}(D\to V(i)\ell^+\nu_{\ell}){\cal A}(V(i)\to P_1P_2P_3){\cal A^*}(D\to V(j)\ell^+\nu_{\ell}){\cal A^*}(V(j)\to P_1P_2P_3)\,.\nonumber
\end{eqnarray}
Then, one can easily obtain
\begin{eqnarray}
  \int_{S_{3}}d\phi_1 \sum_{i=0,\pm}\mid{{\cal A}(D\to V(i)\ell^+\nu_{\ell}){\cal A}(V(i)\to P_1P_2P_3)}\mid^2&=& 0\,.
\end{eqnarray}
and
\begin{eqnarray}
  &&\int_{S_{3}}d\phi_1 \mid{\sum_{i=0,\pm}{\cal A}(D\to V(i)\ell^+\nu_{\ell}){\cal A}(V(i)\to P_1P_2P_3)}\mid^2\nonumber\\
  &=& \int_{S_{3}}d\phi_1\sum_{i\neq j}{\cal A}(D\to V(i)\ell^+\nu_{\ell}){\cal A}(V(i)\to P_1P_2P_3)\nonumber\\
  &\times&{\cal A^*}(D\to V(j)\ell^+\nu_{\ell}){\cal A^*}(V(j)\to P_1P_2P_3)\,.
\end{eqnarray}
With the information of Eq.~\ref{eq:coherent}, the factor $\int_{S_{3}}d\phi_1$
is a coherent term with different polarization for vector mesons and cannot be
measured within the three-body decay formalism. All the three angular
asymmetries $\Upsilon_1$, $\Upsilon_2$ and $\Upsilon_3$ incorporate the factor
of $\int_{S_{3}}d\phi_1$ and are distinctive characteristics in five-body
decays, offering additional constraints on the form factors of $D\to V$
transitions.
  
\begin{figure}[htbp]
  \includegraphics[width=2.0in]{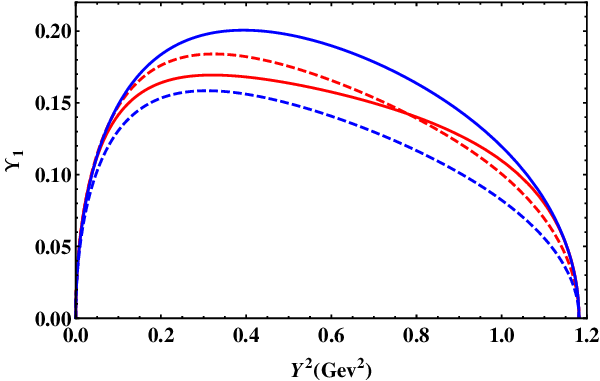}
  \includegraphics[width=2.0in]{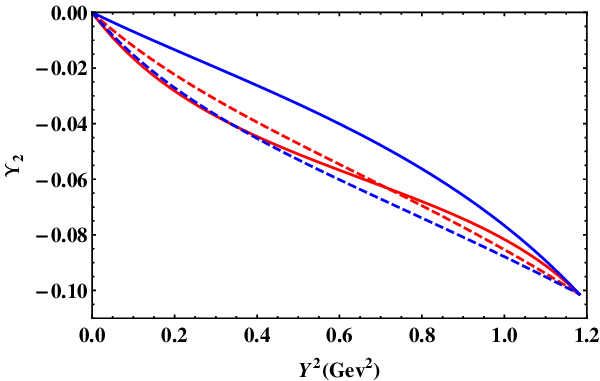}
  \includegraphics[width=2.0in]{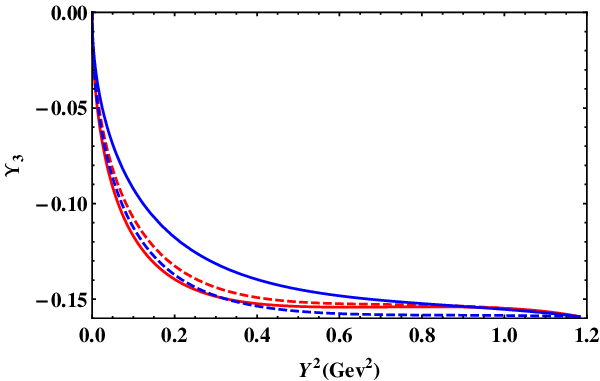}
  \caption{ The distributions of $\Upsilon_1$, $\Upsilon_2$, and $\Upsilon_3$
    related to $Y^2$. The red-solid, red-dashed, blue-solid, blue-dashed lines
    indicate the parametrizations from R.~N.~Faustov
    {\it et al.}~\cite{Faustov:2019mqr}, R.~C.~Verma~\cite{Verma:2011yw},
    S.~Fajfer {\it et al.}~\cite{Fajfer:2005ug}, and M.~A.~Ivanov
    {\it et al.}~\cite{Ivanov:2019nqd},
    respectively.}
  \label{triangle2}
\end{figure}

In the numerical analysis, we test four parametrizations of the $D\to \omega$
transition form factors, given by
\begin{itemize}
\item R.~N.~Faustov {\it et al.}~\cite{Faustov:2019mqr}
  \begin{eqnarray}
    V_0(A_0) &=& \frac{f_0}{(1-Y^2/M^2)(1-a Y^2/M^2+b Y^4/M^4)}\,,\nonumber\\
    A_{1,2} &=& \frac{f_0}{1-a Y^2/M^2+b Y^4/M^4}\,.
  \end{eqnarray}
\item R.~C.~Verma~\cite{Verma:2011yw}
  \begin{eqnarray}
    V_{0}(A_{0,1}) &=& \frac{f_0}{1-a Y^2/M^2+b Y^4/M^4}\,,\nonumber\\
    A_2&=& \frac{f_0}{(1-Y^2/M^2)(1-a Y^2/M^2+b Y^4/M^4)}\,.
  \end{eqnarray}
\item S.~Fajfer {\it et al.}~\cite{Fajfer:2005ug}
  \begin{eqnarray}
    V_0(A_0)&=& \frac{f_0}{(1-Y^2/M^2)(1-a Y^2/M^2)}\,,\nonumber\\
    A_1&=& \frac{f_0}{1-a Y^2/M^2}\,,\nonumber\\
    A_{2}&=& \frac{f_0}{(1-a Y^2/M^2)(1-b Y^2/M^2)}\,.
  \end{eqnarray}
\item M.~A.~Ivanov {\it et al.}~\cite{Ivanov:2019nqd}
  \begin{eqnarray}
    A_0 &=& \frac{M_A-M_V}{2M_V}\left[ A_{0C}-A_2-\frac{Y^2}{M^2_A-M^2_V}A_-\right]\,,\nonumber\\
    A_1 &=& \frac{M_A-M_V}{M_A+M_V}A_{0C}\,,\nonumber\\
    V_0(A_{0C,-,2})&=& \frac{f_0}{1-a Y^2/M^2+b Y^4/M^4}\,.
  \end{eqnarray}
\end{itemize}
Here, $M$ is the pole mass and $f_0$ represents the constant specific to each
form factor at $Y^2=0$. The inputs are summarized in Table~\ref{tab2}. The
distributions of $\Upsilon_1$, $\Upsilon_2$, and $\Upsilon_3$ related to $Y^2$
with different parametrizations of the $D\to \omega$ transition form factors
are shown in Fig.~\ref{triangle2}. One can notice that the model from
S.~Fajfer~{\it et al.}~\cite{Fajfer:2005ug} has the highest asymmetry and
various models demonstrate distinct behaviors in these distributions.
\begin{table}[!bthp]
  \caption{The inputs for the $D\to \omega$ transition form factors. ``-''
    indicates not available.}\label{tab2}
  \begin{tabularx}{\textwidth}{|c|XXX|XXX|XXX|XXX|}
    \hline
    &\multicolumn{3}{c}{Faustov~\cite{Faustov:2019mqr}}&\multicolumn{3}{|c|}{Verma~\cite{Verma:2011yw}}&\multicolumn{3}{|c|}{Fajfer~\cite{Fajfer:2005ug}}&\multicolumn{3}{|c|}{Ivanov~\cite{Ivanov:2019nqd}}\\
    \hline
    & $f_0$ &$a$ &$b$ &$f_0$ &$a$ &$b$ & $f_0$ &$a$ &$b$ &$f_0$ &$a$ &$b$\\
    $V_{0}$ & $0.871$ & $0.146$ &$-2.775$ &$0.85$ &$1.24$ &$0.45$ & $1.05$ & $0.55$ & - &$0.72$ &$1.10$ &$0.27$\\
    $A_{0}[A_{0C}]$ & $0.647$ & $0.224$ &$-0.759$ &$0.64$ &$1.08$ & $1.50$ & $1.32$ & $0.52$ & - &$1.41$ &$0.53$ & $-0.10$\\
    $A_1[A_-]$ & $0.674$ &$0.542$ &$0.350$ &$0.58$ &$0.49$ & $0.02$ & $0.61$ &$0.69$ & - &$-0.69$ &$1.17$ & $0.27$\\
    $A_2$ & $0.713$ & $0.997$ &$2.176$ &$0.49$ &$0.95$ & $0.28$ & $0.31$ & $0.69$ &$0.00$ &$0.55$ &$1.01$ & $0.17$\\
    \hline
  \end{tabularx}
\end{table}

\section{Summary}
Semileptonic decays provide an ideal environment for studying the dynamic
behavior of the strong interaction during the decay process due to the
isolation the hadronic and weak currents from each other. BESIII is currently
accumulating data samples with an integrated luminosity of 20~fb$^{-1}$ at
center-of-mass energy 3.773~GeV~(for $D^0$ and $D^\pm$
mesons)~\cite{BESIII:2024lbn} and has collected 7.33~fb$^{-1}$ of data samples
at $4.128-4.226$~GeV~(for $D_s^+$ mesons)~\cite{BESIII:2020nme}. These datasets
will significantly enhance our sensitivity to investigate the dynamics of
five-body semileptonic decays in the charm sector, which are relatively rare
compared to three- and four-body decays.

We have first derived the general form of the phase space for five-body decays,
and subsequently conducted a systematic investigation of the decay rate and
angular distributions for the decay cascade
$D\to V\ell^+\nu_{\ell}\to(P_1P_2P_3)\ell^+\nu_{\ell}\,(\ell=e,\mu)$. In this
derivation, the mass of electron/muon is explicitly included. With three
additional degrees of freedom compared to four-body decays, five-body decays
offer valuable insights into the decay mechanism. As an example, we have
examined the decay $D+\to\omega e^+\nu_e\to\pi^+\pi^-\pi^0e^+\nu_e$ using
the $D\to\omega$ form factors from various
models~\cite{Faustov:2019mqr, Verma:2011yw, Fajfer:2005ug, Ivanov:2019nqd}.

It is important to note that non-resonant hadronic matrix elements is given as
\begin{eqnarray}
  \langle P_1P_2P_3\mid V^{\mu}(A^{\mu})\mid D\rangle\ &=& f_1p_1^{\mu}+f_2p_2^{\mu}+f_3p_3^{\mu}+f_4q^{\mu}\,,\nonumber\\
  &&+ig_1\epsilon^{\mu\nu\alpha\beta}q_{\nu}p_{1\alpha} p_{2\beta}+ig_2\epsilon^{\mu\nu\alpha\beta}q_{\nu}p_{1\alpha} p_{3\beta}\,,
\end{eqnarray}
where $q=p_V-p_1-p_2-p_3$, the form factors $f_{1,2,3,4}$ and $g_{1,2}$ are the
functions of $p_1\cdot p_2$, $p_1\cdot p_3$, $p_2\cdot p_3$, $p_1\cdot q$,
$p_2\cdot q$, and $p_3\cdot q$. There are six form factors and they are not
simply functions of a single Lorentz invariant. Therefore, it is difficult to
consider the contributions of non-resonant states in general situations.
Future experimental studies should exam the significance of the contribution
from non-resonance amplitudes before proceeding to measure the $D\to V$ form
factors and relevant physics quantities.

Our calculation provides a general method for handling the phase space of
five-body decay and presents the corresponding angular distributions for
semileptonic decays, offering guidance for more complete extraction of form
factors and other parameters in experiments.

\section*{ACKNOWLEDGMENTS}
The authors thank Dr.~Chia-Wei Liu for helpful discussions.
Yechun Yu, Han Zhang and Bai-Cian Ke were supported in part by the Excellent Youth Foundation of Henan Scientific Committee under Contract No.~242300421044, National Natural Science Foundation of China~(NSFC) under Contracts No.~11875054 and No.~12192263, and Joint Large-Scale Scientific Facility Fund of the NSFC and the Chinese Academy of Sciences under Contract No.~U2032104. Yao Yu and Zhuang Xiong was supported in part by NSFC under Contracts No.~11905023, No.~12047564 and No.~12147102, the Natural Science Foundation of Chongqing (CQCSTC) under Contract No.~cstc2020jcyj-msxmX0555, and the Science and Technology Research Program of Chongqing Municipal Education Commission (STRPCMEC) under Contracts No.~KJQN202200605 and No.~KJQN202200621; Jia-Wei Zhang was supported by NSFC under Contract No.~12275036, CQCSTC under Contract No.~cstc2021jcyj-msxmX0681, and STRPCMEC under Contract No.~KJQN202001541. Xue-Wen Chen was supported  by CQCSTC under Contract No.~cstc2021jcyj-msxmX0678 and STRPCMEC under Contracts No.~KJQN202201527 .

\end{document}